\documentclass[twocolumn,prb]{revtex4-1}
\date{today}
\usepackage{natbib,hyperref}
\usepackage{graphicx}
\usepackage{epstopdf}
\usepackage{amsmath}
\usepackage[final]{changes}

\newcommand{\ket}[1]{\left|#1\right>}
\newcommand{\bra}[1]{\left< #1 \right|}

\newcommand{\beq}{\begin{equation}}
\newcommand{\eeq}{\end{equation}}
\newcommand{\beqa}{\begin{eqnarray}}
\newcommand{\eeqa}{\end{eqnarray}}

\begin{document}
\title{Impact of $\mathcal{T}$-symmetry on spin decoherence and control in a synthetic spin-orbit field}
\author{Peihao Huang$^1$}
\email{huangph@sustech.edu.cn}
\author{Xuedong Hu$^2$}
\affiliation{$^1$Shenzhen Institute for Quantum Science and Engineering, Southern University of Science and Technology, Shenzhen 518055, China\\
$^2$Department of Physics, University at Buffalo, SUNY, Buffalo, New York 14260}


\date{\today}

\begin{abstract}
The electrical control of a spin qubit in a quantum dot relies on spin-orbit coupling (SOC), which could be either intrinsic to the underlying crystal lattice or heterostructure, or extrinsic via, for example, a micro-magnet.
Here we show that a key difference between the intrinsic SOC and the synthetic SOC introduced by a micro-magnet is their symmetry under time reversal. Specifically, the time-reversal symmetry ($\mathcal{T}$-symmetry) of the intrinsic SOC leads to not only the traditional van Vleck cancellation known for spin relaxation, but also vanishing spin dephasing to the lowest order of SOC, which we term as ``longitudinal spin-orbit field cancellation''.  On the other hand, the synthetic SOC from a micro-magnet breaks the $\mathcal{T}$-symmetry, therefore eliminates both the ``van Vleck cancellation'' and the ``longitudinal spin-orbit field cancellation''. In other words, the effective field $\vec\Omega$ experienced by the spin qubit does not depend on the quantization magnetic field anymore, and a longitudinal component is allowed for $\vec\Omega$ to the first order of SOC.
Consequently, spin relaxation and dephasing are qualitatively modified compared with the case of the intrinsic SOC. Furthermore, the fidelity of electric-dipole spin resonance based on $\vec\Omega$ could be optimized, with potential applications in spin-based quantum computing.

\end{abstract}

\maketitle

\section{Introduction}
An electron spin qubit in a semiconductor quantum dot (QD) is a promising candidate for quantum information processing due to the long spin coherence time and possible scalability \cite{loss1998, petta2005, hanson_spins_2007, morton_embracing_2011, zwanenburg_silicon_2013, kloeffel_prospects_2013}.
Exciting progress has been made in recent years on spin qubits, such as
high fidelity spin manipulation in a Si QD \cite{yoneda2018, leon_coherent_2020},
strong coupling between a spin qubit in a Si/SiGe QD and a superconducting resonator \cite{mi_coherent_2018, samkharadze_strong_2018, landig_coherent_2018},
two-qubit CPHASE and CNOT gates based on the exchange interaction \cite{veldhorst2015, zajac_resonantly_2018, watson_programmable_2018}, and \added{high fidelity two-qubit gates at an elevated 1.5 K temperature for potential integration of classical and quantum electronics \cite{yang_operation_2020, petit_universal_2020}}.
Indeed, up to nine controllable spin qubits in QDs have been demonstrated \cite{zajac_reconfigurable_2015, mills_shuttling_2018}.

A driving force behind many of the experimental achievements is the introduction of a micromagnet next to the QDs.  The micromagnet creates an inhomogeneous magnetic field, which acts as a synthetic spin-orbit coupling (s-SOC) and allows electrical control of the spin qubit, leading to fast electric dipole spin resonance (EDSR) \cite{tokura_coherent_2006, pioro-ladriere_micromagnets_2007, pioro-ladriere_electrically_2008, kawakami_electrical_2014, wu_two-axis_2014, yoneda2018} and strong spin-photon coupling \cite{hu_strong_2012, mi_coherent_2018, samkharadze_strong_2018, benito_optimized_2019}.
However, while the driving electric field could be applied \added{externally or from a superconducting resonator} \cite{tokura_coherent_2006, pioro-ladriere_micromagnets_2007, pioro-ladriere_electrically_2008, kawakami_electrical_2014, wu_two-axis_2014, yoneda2018, hu_strong_2012, mi_coherent_2018, samkharadze_strong_2018}, it could also be from electrical noises \cite{yoneda2018, borjans_single-spin_2019, hollmann_large_2020, struck_low-frequency_2020, zhang_giant_2020}. 
In other words, the micromagnet and the associated s-SOC open new spin decoherence channels. Various aspects of s-SOC-enabled decoherence have been explored previously, such as spin relaxation \cite{tokura_coherent_2006, borjans_single-spin_2019, hollmann_large_2020, zhang_giant_2020} and dephasing \cite{yoneda2018, struck_low-frequency_2020, li_spin_2019}, and effects of the magnetic noise from the micro-magnets \cite{neumann_simulation_2015, kha_micromagnets_2015}.
However, there is still a lack of understanding on how exactly the s-SOC differs from the intrinsic SOC (i-SOC) in principle, and how such difference affects both spin coherence and control in QD systems.

In this work, we show that the most important qualitative difference between s-SOC and i-SOC is with respect to the time-reversal symmetry ($\mathcal{T}$-symmetry): s-SOC breaks it, while i-SOC preserves it.
For the case of the i-SOC, we show that it leads to the traditional van Vleck cancellation \cite{van_vleck_paramagnetic_1940, orbach_theory_1961}.  More importantly for a spin qubit, it also results in vanishing longitudinal component of the effective magnetic field, which we term as ``longitudinal spin-orbit field cancellation'', so that electrical fluctuations (whether from phonons or background charge fluctuations) cannot lead to pure spin dephasing. 
For s-SOC, however, the broken $\mathcal{T}$-symmetry leads to the disappearance of both these cancellations.
Consequently, spin relaxation has a different magnetic field dependence in the presence of s-SOC as compared to i-SOC, and longitudinal effective magnetic field is allowed at the lowest order of s-SOC, so that electrical noise could cause non-negligible pure dephasing of a spin qubit.  In addition, we show that the fidelity of the s-SOC induced EDSR could be improved through proper orientation of the applied magnetic field. 
It is important to emphasize here that the connection between spin qubit properties and $\mathcal{T}$-symmetry of the SOC is general, and is present in many other physical systems, so that the studies we perform here should be relevant in those situations.

\section{System Hamiltonian}
\begin{figure}[]
\includegraphics[scale=0.35]{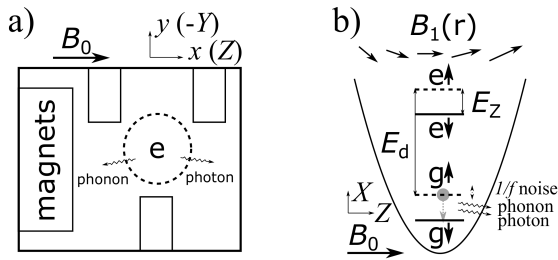}
\caption{Schematic diagrams. (a) Schematics of an electron spin qubit in a gate-defined QD next to a micromagnet.
\added{An external magnetic field is applied along the $x$-axis and polarizes the micromagnet. 
$Z$-axis is defined along the spin quantization direction.
Spin relaxation happens in the presence of s-SOC from the inhomogeneous magnetic field and electrical noises, such as $1/f$ charge noise, phonons, or photon fluctuations. 
(b) Schematics of the energy levels in a QD. 
The parabolic curve represents the confinement potential.
The two solid lines are orbital states, and two dashed lines represents the lifted spin degeneracies in a uniform field $B_0$. 
The arrows on the top represents the inhomogeneous field $\vec{B}_1$ from the micromagnet, which has both the transverse and longitudinal field gradients. 
The s-SOC from $\vec{B}_1$ hybridizes the spin states and results in the spin-electric-field coupling.}
}\label{schematics}
\end{figure}

We consider an electron in a single gate-defined QD [see Figure \ref{schematics} (a)].
The QD confinement in the [001]-direction (defined as $z$-axis) is provided by the interface electric field, while the in-plane (i.e. $xy$ plane) confinement is provided by top gates.  A micromagnet (e.g. a cobalt magnet) is deposited over the QD and polarized by an applied magnetic field.
We separate the total magnetic field into two parts $\vec{B}=\vec{B}_0+\vec{B}_1$, where $\vec{B}_0$ ($\vec{B}_1$) is the position-independent (position-dependent inhomogeneous) magnetic field. The Hamiltonian for this model system is thus
\begin{equation}
H=H_Z + H_d + H_{SO} +  V_{ext}(\vec{r}).
\end{equation}
Here $H_Z= \frac{1}{2}g \mu_B \vec{\sigma} \cdot \vec{B}_0$ is the Zeeman Hamiltonian due to the position independent field, where $g$ is the effective g-factor, and $\vec{\sigma}$ is the Pauli operator for the electron spin.  $H_{d} =\frac{p^2}{2m^*} + \frac{1}{2}m^* \omega_d^2 r^2$ is the usual electron 2D orbital Hamiltonian in a single QD, where $\vec{r}=(x,y)$ and $\vec{p}=-i\hbar\nabla + (e/c)\vec{A}(\vec{r})$ are the in-plane 2D coordinate and kinetic momentum operators ($e>0$), and $\omega_d$ is the characteristic frequency of the in-plane confinement. The out-of-plane dynamics is neglected due to the strong confinement at the interface. $V_{ext}(\vec{r})$ is the external electric potential, from sources such as electrical noises or a manipulation field.
Lastly, the s-SOC term $H_{SO}$ is due to the Zeeman effect of the position-dependent inhomogeneous magnetic field $\vec{B}_1$.  Keeping the lowest order position dependence,
\beqa 
H_{SO} &=& \frac{1}{2}g \mu_B \vec{\sigma}\cdot \vec{b}_{1}x, \label{Hs-SOC}
\eeqa
where the inhomogeneity is assumed in the $x$-direction without loss of generality, and the gradient of the inhomogeneous magnetic field,  $\vec{b}_{1}\equiv \left.\partial \vec{B}_1/\partial x \right|_{x=0} \equiv [b_{1l},0,b_{1t}]$, is the coupling constant of the s-SOC.
%
%

\section{Effective Hamiltonian}

In general, the energy scale of SOC is much smaller than the orbital and Zeeman energy in a QD.  We can thus perform a Schrieffer-Wolff (SW) transformation $H_{eff}= e^S H e^{-S}$ to eliminate $H_{SO}$ in the leading order by requiring that $[H_d + H_{Z},S] = H_{SO}$ \cite{golovach2004, huang2014a}.  The transformation generator $S$ can be formally written as
\beq \label{Ssum}
S=\sum_{m=0}^{\infty} \left({L_Z}{L_d^{-1}}\right)^m L_d^{-1} H_{SO},
\eeq
where the super-operators $L_d$ and $L_Z$ are defined such that $L_d \mathcal{O}=[H_d,\mathcal{O}]$ and $L_Z \mathcal{O}=[H_{Z},\mathcal{O}]$ for any given operator $\mathcal{O}$. Once $S$ is given, an effective spin Hamiltonian $H_{eff}=H_Z+[S,V_{ext}]$ can be obtained.

\textbf{The critical role of $\mathcal{T}$-symmetry}---
The $\mathcal{T}$-symmetry of $H_{SO}$ plays a critical role in the SW transformation. For any SOC, the generator $S$ can be formally rewritten as $S=\vec{f} \cdot \vec{\sigma}$, where $\vec{f}$ contains the orbital operators. Since both the ground orbital state $\ket{\psi}$ and $V_{ext}(\vec{r})$ are time-reversal symmetric (TRS), the matrix element $\bra{\psi} [\vec{f}, V_{ext}(\vec{r})]\ket{\psi}$ would be finite only when $\vec{f}$ is also TRS. 
Given that spin operator $\vec{\sigma}$ is time-reversal asymmetric (TRA), only the TRA terms in $S$ could contribute to the effective spin Hamiltonian $H_{eff}$.
%
%

For the TRA s-SOC, the first term $L_d^{-1} H_{SO}$ in $S$ is TRA (see Appendix A), which is allowed. For the TRS i-SOC, on the other hand, the first term in $S$ is TRS (see Appendix A), so that it is forbidden. 
The lowest order contribution would come from the next order term $L_Z L_d^{-2} H_{i-SOC}$, which is TRA and is allowed, with $L_Z$ involving Zeeman term $H_Z$ that breaks $\mathcal{T}$-symmetry. 
%
Clearly, the different $\mathcal{T}$-symmetries of the s-SOC and the i-SOC ensures that their contribution to the spin Hamiltonian are of different orders in $E_Z/E_d$, and leads to qualitatively different results.

More specifically, for i-SOC, the lowest-order contribution is from the second term in $S$, which is linearly proportional to $B_0$.  This leads to an extra $B_0^2$ dependence in the spin relaxation rate (shown below), underlying the so-called van Vleck cancellation \cite{van_vleck_paramagnetic_1940, orbach_theory_1961}. Moreover, given that all the higher-order terms in $S$ contains the $L_Z$ operator, and the property $L_Z (\vec{\varepsilon}\cdot \vec\sigma) \propto (\vec{B}_0\times \vec{\varepsilon})\cdot \vec{\sigma}$ is satisfied by any vector $\vec{\varepsilon}$, so that the resulting effective spin-electric coupling $[S,V_{ext}]\propto (\vec{B}_0\times [\vec{\varepsilon},V_{ext}])\cdot \vec\sigma$ contains only transverse coupling, and the effective field sensed by the spin is always transverse.  In short, the conservation of the $\mathcal{T}$-symmetry of the i-SOC results in the vanishing longitudinal effective magnetic field to any order in the perturbative expansion of Eq.~(\ref{Ssum}), and spin dephasing via SOC with $\mathcal{T}$-symmetry would vanish to the first order of SOC and all orders of Zeeman interaction. We term this effect ``longitudinal spin-orbit field cancellation'', in analogy to the van Vleck cancellation for spin relaxation. In comparison, the breaking of $\mathcal{T}$-symmetry by the s-SOC means that generator $S$ could contain a term $L_d^{-1} H_{SO}$ that is independent of $L_Z$ operator, therefore removing the condition for both the van Vleck cancellation and the ``longitudinal spin-orbit field cancellation''. In other words, for s-SOC, the effective magnetic field could be independent of $B_0$, and the longitudinal effective magnetic field is allowed.

Our analysis here does not assume any specific form of the SOC and $H_d$ other than their $\mathcal{T}$-symmetry.  It is thus generally applicable to any physical system as long as the SOC is smaller in magnitude than the orbital and Zeeman splitting. For example, the results can be used for an electron spin qubit in a double QD  with or without a micro-magnet, or an electron spin at donor(s) with a Coulombic confinement. Moreover, the theory could also be applicable to other forms of SOC. 

\textbf{Effective Hamiltonian}---
\added{We consider the case that orbital splitting is large, so that} $||H_{s-SOC}|| \ll E_Z =g\mu_B B_0 \ll E_d=\hbar\omega_d$, which is the general condition underlying a spin qubit in a QD. To the lowest order of $H_{s-SOC}$ and $E_Z/E_d$, $S \approx L_d^{-1} H_{SOC}$.
Using this generator, the effective spin Hamiltonian is (see Appendix A) 
\beq
H_{eff}=\frac{1}{2}g\mu_B \vec{\sigma}\cdot \left(\vec{B}_0 + \vec{\Omega}\right),
\eeq
where $\vec{\Omega}=-\vec{b}_{1}\partial_x V_{ext}/(m^*\omega_d^2)$ is the effective magnetic field generated from the electric potential through the QD displacement $\delta_x= -\partial_x V_{ext}/(m^*\omega_d^2)$ and the s-SOC. 

As we have discussed in the previous section, the effective field $\vec{\Omega}$ generated by $S = L_d^{-1} H_{SOC}$ has some distinctive properties as a result of the elimination of both the van Vleck cancellation and the longitudinal spin-orbit field cancellation. First, it can have a longitudinal component parallel to the constant applied magnetic field $\vec{B}_0$. If $V_{ext}$ is a potential from electrical noise, the corresponding effective magnetic noise $\vec\Omega$ would in general lead to both spin relaxation and pure dephasing (see Figure \ref{schematics} (b)).  Second, $\vec\Omega$ is independent of $B_0$.  Both are different from the case of the i-SOC, where the effective noise magnitude is linearly proportional to $B_0$, and induces only spin relaxation \cite{golovach2004, huang2014a}.
%
Below we explore consequences of these new characteristics of the effective field $\vec\Omega$.


\section{Spin Relaxation}
Spin relaxation time gives the upper limit of spin coherence time.
Suppose the direction of magnetic field $\vec{B}_0$ (assumed along the $x$-axis in this work) is defined as the new $Z$-axis, while $X$- and $Y$-axis are orthogonal to the $Z$-axis (see Figure \ref{schematics} (b)), the relaxation rate is then given by
$
1/T_1= S_{XX}(\omega_Z) + S_{YY} (\omega_Z)
$
\cite{golovach2004, huang2014a}, where $S_{ii}(\omega)$ is the power spectral density of the magnetic noise in the $i$th direction, and $\omega_Z = g\mu_BB_0/\hbar$ is the Larmor frequency. 
In other words, spin relaxation is determined by the transverse magnetic noise $\Omega_X=-{b}_{1t}\partial_x V_{ext}/(m^*\omega_d^2)$, 
and the relaxation rate is
\beq
1/T_1 =  \left[\frac{g\mu_B {b}_{1t}}{2\hbar m^*\omega_d^2}\right]^2 S_{FF} (\omega_Z),
\eeq
where $S_{FF} (\omega)$ is the spectral density of the force correlation of the noise (see Appendix B) \cite{huang2014b}.
\added{When $E_Z>E_d$, there could be an additional spin relaxation channel via an intermediate state, which flips the spin and orbital states simultaneously, followed by a direct orbital relaxation \cite{yang2013, srinivasa_simultaneous_2013}. However, in the limit of $E_Z\ll E_d$, as assumed in deriving our effective Hamiltonian, this spin relaxation mechanism is absent.
}

Qualitatively, spin relaxation depends on the transverse field gradient, with $1/T_1 \propto b_{1t}^2=(\partial{B_z}/\partial x)^2$.
It also has a strong dependence on the QD confinement, $1/T_1 \propto 1/\omega_d^4$, and is thus suppressed in a smaller QD.
Lastly, the dependence of spin relaxation on the magnetic field $B_0$ is given by $S_{FF} (\omega_Z)$ (neglecting the weak dependence of $b_{1t}$ on $B_0$ when the micromagnet is saturated), as we discuss in more detail below.

\begin{figure}[]
\includegraphics[scale=0.35]{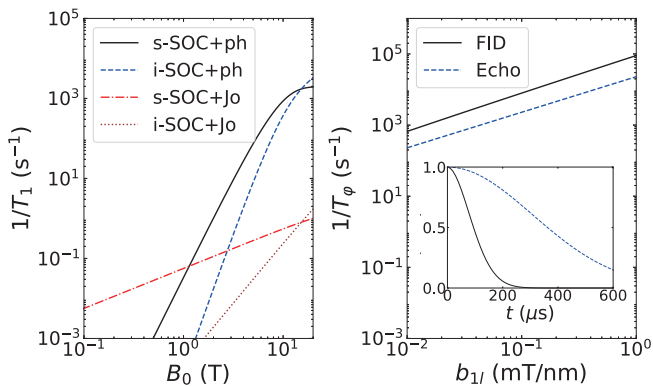}
\caption{
\added{Spin relaxation and pure dephasing. (a) Spin relaxation $1/T_1$ mediated by the s-SOC or the i-SOC as a function of the magnetic field $B_0$ due to deformation phonon and Johnson noise. 
The spin relaxation mediated by the s-SOC has weaker $B_0$ dependence than the i-SOC due to the broken $\mathcal{T}$-symmetry of the s-SOC.
(b) Spin pure dephasing $1/T_\varphi$ and $1/T_{\varphi,echo}$ (Hahn echo) mediated by the s-SOC due to $1/f$ charge noise as a function of the longitudinal field gradient $b_{1l}$.
The pure dephasing rates increase as the longitudinal field gradient $b_{1l}$ increases.
Inset of (b): Spin dephasing dynamics for free induction decay or Hahn echo ($b_{1l}=0.1$ mT/nm).}
}\label{combined}
\end{figure}

Figure \ref{combined}(a) shows the spin relaxation rate as a function of the magnetic field $B_0$ due to deformation coupling to phonons or dipole coupling to Johnson noise in silicon (parameters are listed in Appendix A). 
Here we neglect the valley states (from conduction band degeneracy) in silicon and focus on the intra-valley spin-orbit mixing. 
Spin relaxation due to phonon emission mediated by s-SOC shows a $B_0^5$ dependence, $1/T_1\propto B_0^5$, in contrast to the $B_0^7$ dependence in the case of the i-SOC \cite{khaetskii2001,golovach2004}, where the extra $B_0^2$ dependence is due to the $\mathcal{T}$-symmetry of the i-SOC that leads to the Van Vleck cancellation.
For the same reason, spin relaxation due to Johnson noise mediated by s-SOC shows a linear $B_0$ dependence, $1/T_1 \propto B_0$, in contrast to the $B_0^3$ dependence in the case of the i-SOC \cite{huang2014a}.
%
%
Consequently, at low $B$-field spin relaxation is dominated by Johnson noise, while at higher $B$-field it is dominated by phonon emission, as shown in Figure \ref{combined}.
Numerically, spin relaxation rate grows from $0.01$ s$^{-1}$ to 1000 s$^{-1}$ as $B_0$ increases from $0.2$ T to 10 T.

The spin relaxation channel discussed here can be relevant in a silicon QD experiment considering that i-SOC is generally weak in Si. For comparison, let us consider the case of an electron spin in a silicon QD without a micromagnet. The spin relaxation is mostly due to phonon and Johnson noise via the i-SOC induced spin-valley or spin-orbit mixing \cite{huang2014b}: At low magnetic field, spin relaxation is dominated by the spin-valley mixing, where $1/T_1 \sim 1$ to $1000$ $s^{-1}$. 
When Zeeman splitting is less than the orbital splitting but larger than the valley splitting (i.e. $B_0>B_{VS}=E_{VS}/(g\mu_B)\sim 3$ T if the valley splitting $E_{VS}$ is 0.2 meV), spin relaxation via spin-valley mixing is strongly suppressed, and spin relaxation is dominated by spin-orbit mixing [corresponding to the case of i-SOC+ph in Fig.~\ref{combined} (a)].
The numerical estimate in Fig.~\ref{combined}(a) thus indicates that relaxation due to phonon noise via s-SOC can be dominant in a silicon QD when $E_{VS}<E_Z\ll E_d$, where the spin relaxation shows $B_0^5$ dependence. Recent experiments in a single Si QD with micromagnets show that spin relaxation in the high field limit has a $B_0^5$ dependence \cite{borjans_single-spin_2019, hollmann_large_2020}, deviating from the $B_0^7$ dependence normally observed in silicon without micromagnets \cite{amasha_electrical_2008, hayes_lifetime_2009, xiao_measurement_2010}, but consistent with our theoretical results here.

\added{
At magnetic fields when $B_0<B_{VS}$, experimental results also show the difference between the s-SOC and i-SOC \cite{borjans_single-spin_2019, hollmann_large_2020}, consistent with our theory. However, due to the generally small valley splitting in these experiments, valley effects have to be included \cite{huang2014b} to explicitly show the applicability of our result. 
A recent study indicates that the results, such as the $B_0$ and $b_{1t}$ dependences of spin relaxation, due to the $\mathcal{T}$-symmetry property of SOC is still valid for spin-valley mixing \cite{huang_fast_2020}. 
%
For simplicity, we focus on the spin-orbital mixing in this study, although the results based on the symmetry argument are generally applicable to other cases such as spin-valley mixing, or a spin qubit in a double QD.
%
Moreover, 
in a device with small orbital excitation energy (large QD) and large valley splitting, $E_d <E_{VS}$, where the spin-valley relaxation is suppressed and the intra-valley spin-orbit mixing is dominant, our results is relevant at low $B$ fields. 
}


\section{Pure dephasing}
The breaking of $\mathcal{T}$-symmetry by s-SOC allows a longitudinal effective magnetic field, so that electrical noise can cause pure dephasing of the electron spin at the lowest order of s-SOC, which is forbidden with i-SOC \cite{golovach2004,huang2014a}. Such charge noise induced dephasing has indeed been measured and discussed in Refs.~\cite{yoneda2018, struck_low-frequency_2020}, where it has been demonstrated that the $1/f$ charge noise induced spin pure dephasing is the dominant dephasing channel in the presence of s-SOC in their experiments. Below we examine the qualitative dependence of the spin pure dephasing on system parameters such as the QD size and the field gradient, and give some numerical results.

Pure dephasing is determined by the spectral density $S_{ZZ}(\omega)$ of the effective longitudinal magnetic noise \cite{hu2006, huang_spinDephasing_2018}.  Suppose the spectral density of the electrical potential fluctuation of $1/f$ charge noise is $S_{1/f}=A/\omega$, where $A$ characterizes the noise strength. For the spin qubit, we then have $S_{ZZ}(\omega) = A_{eff}/ \omega$, where $\sqrt{A_{eff}} = \sqrt{A} \left[g\mu_B {b}_{1l}/(2\hbar m^*\omega_d^2 l_0) \right]$ is the amplitude of the effective magnetic noise, and the length $l_0$ converts the electric potential to the field strength of the charge noise.

The magnetic noise amplitude $\sqrt{A_{eff}}$ depends on the longitudinal field gradient and the orbital splitting as $\sqrt{A_{eff}} \propto {b}_{1l} / \omega_d^2$, consistent with the intuitive argument given in the supplementary material in Ref.~\cite{yoneda2018}. We emphasize here that this intuition works only in the case for the s-SOC, where the $\mathcal{T}$-symmetry is broken, but does not work in the case for the i-SOC \cite{huang2014a}.  Due to the initial Gaussian time dependence for the off-diagonal density matrix element, the spin pure dephasing rate $1/T_{\varphi} \propto \sqrt{A_{eff}}$ \cite{huang_spinDephasing_2018}.  Thus the pure dephasing rate is determined by the amplitude of the $1/f$ charge noise, the magnitude of the longitudinal field gradient, and QD confinement as
\beq
1/T_{\varphi} \propto \frac{b_{1l} \sqrt {A}}{\omega_d^2} \,.
\eeq

Figure \ref{combined}(b) shows the spin dephasing rate with or without echo as a function of $b_{1l}$ due to $1/f$ charge noise.  The dephasing rate is extracted numerically from the dynamics as shown in the inset (calculated similarly as in Ref. \onlinecite{huang_spinDephasing_2018}).  The spin dephasing rate $1/T_\varphi$ (or $1/T_{\varphi,echo}$) has an approximately linear dependence on the magnetic field gradient $b_{1l}$, consistent with the expectation based on the analytical expressions. Quantitatively, the dephasing rate goes from 8000 s$^{-1}$ to 90000 s$^{-1}$ (or from 2300 s$^{-1}$ to 30000 s$^{-1}$ with spin echo) as the magnetic field gradient $b_{1l}$ increases from 0.1 mT/nm to 1 mT/nm. The pure dephasing rate can vary if the amplitude of charge noise or orbital confinement varies [Here, the dephasing rate at $b_{1l}$=0.5 mT/nm is $1/T_{\varphi}\sim 10^4$ s$^{-1}$ corresponding to the value observed in Ref.~\cite{yoneda2018}].

The pure dephasing due to charge noise can be compared with other possible dephasing mechanisms. For an electron spin in a QD without a micromagnet, nuclear spin noise is a major source for spin dephasing, with spin dephasing time measured to be 360 ns ($1/T_2^* \sim 2.7 \times 10^6$ s$^{-1}$) in natural silicon, and as long as 120 $\mu$s ($1/T_{2}^*\sim 8 \times 10^3$ s$^{-1}$) in isotopically enriched $^{28}$Si \cite{veldhorst2015}. Thus, our numerical estimate indicates that dephasing due to charge noise via s-SOC can be dominant when $b_{1l}$ is bigger than 0.1 mT/nm (also dependent on the charge noise amplitude) in an isotopically enriched silicon QD. The dominance of charge-noise-induced pure dephasing has indeed been observed in recent experiments \cite{yoneda2018, struck_low-frequency_2020}. 

To suppress pure dephasing from charge noise and improve spin coherence in the presence of s-SOC, a straightforward approach is to reduce longitudinal field gradient $b_{1l}$. 
However, \deleted{Maxwell equations for the magnetic field dictate that gradient of the magnetic field along different directions are related and have to be accounted for. }
according to the Maxwell equations $\nabla \cdot\vec{B}=0$ and $\nabla \times \vec{B}=0$. Therefore, in addition to $b_{1l} = \partial B_x/\partial x$ and $b_{1t} = \partial B_z/\partial x$, the gradient $\partial B_z/\partial z$ and $\partial B_x /\partial z$ would have to be finite as well (all other gradients can be zero with translational symmetry along $y$). 
We emphasize that, to optimize the operation of the spin qubit system, we need to consider field gradient and confinement together. 
We neglected the effects of the gradient $\partial B_z/\partial z$ and $\partial B_x /\partial z$ in our case because 
\added{the orbital motion along $z$ axis is suppressed due to the strong vertical confinement.}
Thus, only gradients $b_{1l}$ and $b_{1t}$ have to be taken into account.
In this case, 
if we can adjust the quantization field $\vec{B}_0$, such that $\vec{B}_0$ is perpendicular to the vector $\partial \vec{B}/\partial x = (b_{1l},0,b_{1t})$, then the induced effective magnetic field becomes perpendicular to the quantization axis. Consequently, the pure dephasing from charge noise can be minimized, while the transverse magnetic field is maximized for EDSR. 



%
%


\section{Optimization of spin control using direction of magnetic field}
The discussion above shows that it is possible to optimize spin manipulation by adjusting the direction of the applied magnetic field, which we explore in this section.  

In a conventional spin resonance experiment, a constant magnetic field establishes the quantization axis of the spins, while a small transverse AC magnetic field is used to flip the spins. The same principle is behind EDSR experiments in the context of s-SOC \cite{rashba_orbital_2003, flindt_spin-orbit_2006, golovach2006, nowack_coherent_2007, rashba_theory_2008, tokura_coherent_2006, yoneda2018}. 
%
%
%
%
Here we consider an oscillating electric field $\vec{E}(t)$ that is applied along the $x$-axis, $\vec{E}(t) =\vec{E}_{0}\cos(\omega_0 t + \phi_0)$,
where $\vec{E}_{0}$ is the field magnitude, and $\omega_0$ and $\phi_0$ are the frequency and the phase of the field.
Driven by this electric field, the electron spin experiences an effective oscillating magnetic field via the s-SOC.
The driven spin Hamiltonian takes the form
\beq
H_0=\frac{\omega_Z}{2}\sigma_Z + (\Omega_{0,t}\sigma_X+ \Omega_{0,l} \sigma_Z) \cos(\omega_0 t+\phi_0),
\eeq
where $\Omega_{0,t} = - g\mu_B {b}_{1t}\frac{eE_{k,max}}{2m^*\omega_d^2}$ and $\Omega_{0,l} = - g\mu_B {b}_{1l}\frac{eE_{k,max}}{2m^*\omega_d^2}$ are the maximum transverse and longitudinal magnetic field the spin experiences. 
%
Due to the breaking of $\mathcal{T}$-symmetry by s-SOC, the effective field $\Omega_{0,t}$ is independent of the applied magnetic field, contrary to the i-SOC-mediated EDSR \cite{golovach2006}. Therein lies a significant advantage for s-SOC enabled EDSR: a lower magnetic field can be applied to suppress the spin relaxation without sacrificing the speed of EDSR. Moreover, besides the transverse term $\Omega_{0,t}$, there is in general also a longitudinal term $\Omega_{0,l}$.
%
While the transverse component $\Omega_{0,t}$ is normally used for spin manipulation \cite{tokura_coherent_2006}, below we show that the longitudinal component $\Omega_{0,l}$ can be used to enhance the speed and the fidelity of spin manipulation.

As discussed above, the field gradient along the quantization axis induces spin pure dephasing, which could limit the spin control fidelity, and the choice of quantization axis could reduce the spin dephasing. Here, we explore further the explicit dependence of the Rabi frequency and quality factor of EDSR on the orientation of the spin quantization axis determined by $\vec{B}_0$.
We assume that the micromagnet is fully polarized, so that the field gradient is insensitive to a change of direction of the applied magnetic field. The orientation of $\vec{B}_0$ can be modified by changing the direction of the applied magnetic field \cite{zhang_giant_2020}. To evaluate the spin decoherence, a background dephasing $1/T_{\varphi,0}=10^4$ $s^{-1}$ is assumed from other dephasing mechanisms such as nuclear spins, corresponding to the case of an isotopically purified silicon QD \cite{veldhorst2015}. The total dephasing is then estimated as $1/T_2^* = 1/T_{\varphi,0} + 1/T_\varphi$ [contribution from $T_1$ process can be omitted since it is much slower than pure dephasing at low magnetic fields].

\begin{figure}[]
\includegraphics[scale=0.35]{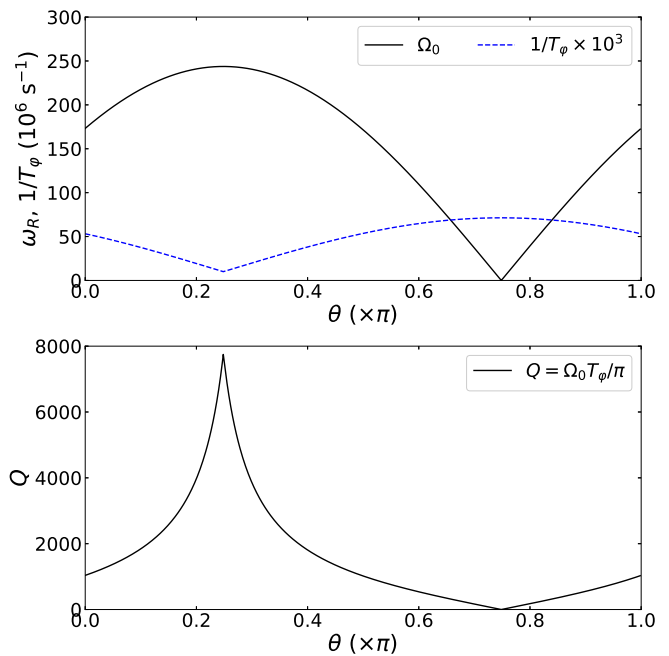}
\caption{
\added{Upper panel: Rabi frequency (black solid) and pure dephasing rate (blue dashed) of a spin qubit versus the polar angle $\theta$ of $\vec{B}_0$ when $b_{1t}=b_{1l}=0.5$ mT/nm. A background dephasing $1/T_{\varphi,0}=10^4$ $s^{-1}$ is assumed from other dephasing mechanisms such as nuclear spins, corresponding to the case of an isotopically purified silicon QD. As the quantization axis goes away from $\theta=0$ (out-of-plane), Rabi frequency is enhanced while the dephasing rate is suppressed. 
Lower panel: Quality factor $Q=\omega_R T_\varphi/\pi$ of the spin control versus the angle $\theta$. Because of the out-of-phase variation of Rabi frequency and suppression of dephasing rate, the quality factor can be enhanced by almost an order of magnitude at the optimal angles.}
}\label{spinManip_thetaOfB}
\end{figure}

As illustrated in Fig.~\ref{spinManip_thetaOfB}, 
as the quantization axis goes away from $\theta=0$ (out-of-plane), Rabi frequency is enhanced, while the pure dephasing rate is suppressed simultaneously.  Because of this change in opposite directions, the quality factor $Q=\omega_R T_\varphi/\pi$ for the spin control can be enhanced by nearly an order of magnitude, as shown in
the lower panel of Fig.~\ref{spinManip_thetaOfB}. 

In short, the breaking of the $\mathcal{T}$-symmetry by the s-SOC leads to quite different behaviors in spin dephasing and EDSR compared with i-SOC. Specifically, EDSR via s-SOC is independent of the magnitude of the applied magnetic field.  Given a particular field gradient vector $\vec{b}_1$, one can optimize EDSR versus pure dephasing by adjusting properly the spin quantization axis.


\section{S-SOC enabled spin-photon coupling}
Spin qubit communication and coupling via a cavity could be a key ingredient in a scalable quantum computer, and have generated extensive theoretical and experimental explorations \cite{imamoglu_quantum_1999, trif_spin-electric_2008, hu_strong_2012, mi_coherent_2018, samkharadze_strong_2018, borjans_resonant_2020}.  With magnetic coupling too weak to be useful, spin-photon coupling via the cavity electric field, assisted by i-SOC or s-SOC, has become the only realistic approach to reach the strong-coupling limit, and has recently been demonstrated experimentally in a Si double QD with a micromagnet \cite{mi_coherent_2018, samkharadze_strong_2018}.  As we have shown above, the breaking of $\mathcal{T}$-symmetry by the s-SOC results in qualitative differences for spin-electric-field coupling compared with the case of i-SOC, which could lead to modifications to the behavior of spin-cavity and spin-spin interaction via a cavity.  Note that to achieve the spin-photon strong coupling, the applied magnetic field should be small so that it does not destroy superconductivity of the cavity electrodes.  When the valley splitting is larger than the orbital splitting in the device (such as a large MOS QD \cite{yang2013}), our theory would be applicable at low fields.  Below, we examine the properties of spin-photon coupling between a spin qubit and a superconducting resonator.

To enhance the cavity electric field, we assume that two gates of the QD are connected to the electrodes of a superconducting resonator \cite{mi_coherent_2018, samkharadze_strong_2018, borjans_resonant_2020}, so that the electric field $E=-\partial_x V_{ext}/|e|$ in the effective spin Hamiltonian is given by the electric field $E_{sc}$ across the QD due to the voltage difference of the center pin and ground of the resonator.  The voltage operator in the resonator is given by $V_{sc}=V_{zpf}(a^\dag + a)$, where $a$ is the photon annihilation operator in the single-mode superconducting resonator. The voltage $V_{zpf}=\omega_0\sqrt{\hbar Z_0}$ is the voltage amplitude due to the zero-point fluctuation (ZPF) in the resonator, where $\omega_0$ is the resonator frequency and $Z_0$ is the characteristic impedance of the resonator. Therefore, the electric field operator becomes $E_{sc}=E_{zpf}(a^\dag + a)$, where $E_{zpf}=V_{zpf}/d_0$ and $d_0$ (about the QD size) is the length for the voltage drop. 
%
%
The spin-photon coupling Hamiltonian thus takes the form
\beq
H_{s-ph}=g_{s,t}(\sigma_+ a + a^\dag \sigma_-) + g_{s,l}\sigma_Z(a + a^\dag),
\eeq
where $\sigma_\pm \equiv \sigma_X \pm i\sigma_Y$ is the spin creation or annihilation operator.
The strength for the transverse and the longitudinal spin-photon coupling are given by
$g_{s,t}= - \frac{eg\mu_B {b}_{1t}}{2m^*\omega_d^2} \omega_0 \sqrt{\hbar Z_0}/d_0,$ and
$g_{s,l}= - \frac{eg\mu_B {b}_{1l}}{2m^*\omega_d^2} \omega_0 \sqrt{\hbar Z_0}/d_0.$
In comparison, for i-SOC and at low magnetic field ($g\mu_BB_0\ll\hbar\omega_d$), $g_{s,l,isoc}=0$, and
$g_{s,t,isoc}= - \frac{eg\mu_B B_0}{\lambda_{so}m^*\omega_d^2d_0} \omega_0 \sqrt{\hbar Z_0},$
where $\lambda_{so}=\hbar/(m^*\alpha)$ is the effective length of i-SOC, with $\alpha$ being the Rashba SOC constant \cite{huang2014a}.

The spin-photon coupling strength $g_{s,t}$ and $g_{s,l}$ have linearly dependence on the resonator frequency $\omega_0$, and $\sqrt{Z_0}$ dependence with the resonator characteristic impedance, and have no dependence on the magnetic field $B_0$. The coupling strengths also have strong dependence on the size of the QD, with $g_{s,t}$, $g_{s,t} \propto 1/(\omega_d^2 d_0) \propto r_{QD}^3$, where $r_{QD}$ is the size of the QD.
Lastly, $g_{s,t}$ ($g_{s,t}$) has linear dependence on the transverse (longitudinal) field gradient, with $g_{s,t}\propto b_{1t}$ ( $g_{s,l}\propto b_{1l}$).

\begin{figure}[]
\includegraphics[scale=0.35]{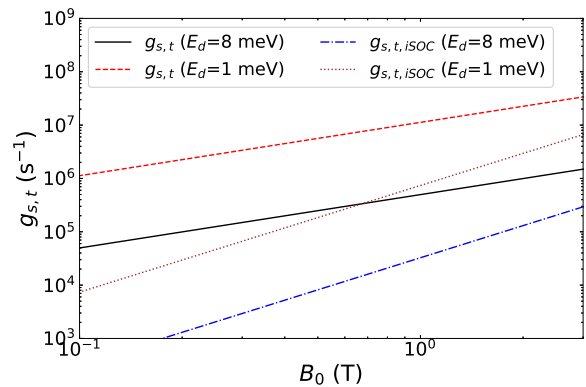}
\caption{Transverse spin-photon coupling strength via s-SOC ($g_{s,t}$) or i-SOC ($g_{s,t,iSOC}$) as a function of the magnetic field $B_0$ for different QD confinement $E_d$. \added{Spin-photon coupling $g_{s,t}$ or $g_{s,t,iSOC}$ increases as $E_d$ decreases. $g_{s,t}$ has weaker $B_0$ dependence compared to $g_{s,t,iSOC}$ due to the broken $\mathcal{T}$-symmetry of the s-SOC.}
}\label{g_s}
\end{figure}

Figure \ref{g_s} shows the spin-photon transverse coupling strength as a function of the magnetic field $B_0$. 
The coupling strength $g_{s,t}$ shows a linear $B_0$ dependence (the spin and the resonator are assumed to be in resonance, i.e. $\omega_0=\omega_Z$), in contrast to the $B_0^2$ dependence for $g_{s,t,isoc}$. The extra $B_0$ dependence is due to the $\mathcal{T}$-symmetry of the i-SOC. 
Indeed, the ratio of the two transverse spin-photon coupling strength satisfies
\beq
\frac{g_{s,t}}{g_{s,t,isoc}} = \frac{b_{1t} \lambda_{so}}{B_0},
\eeq
which is independent of the quantum dot and cavity parameters such as $\omega_d$ and $Z_0$.
With $b_{1t}$ fixed by the fabrication and $\lambda_{so}$ fixed by the material and the interface electric field, $B_0$ becomes an important indicator of the coupling strengths.  Specifically, $g_{s,t}$ can be much stronger than $g_{s,t,isoc}$ at lower magnetic fields.
For example, when $E_d=8$ meV, $g_{s,t}$ is about $5\times10^4$ s$^{-1}$ at $B_0=0.1$ T, and increases linearly as $B_0$ increases, while $g_{s,t,isoc}$ grows from $3\times10^2$ s$^{-1}$ to $3\times10^5$ s$^{-1}$ as $B_0$ increases from $0.1$ T to 3 T. The result suggests that spin-photon strong coupling is possible with s-SOC at low magnetic fields even in a single QD, which is an important consideration when integrating a spin system with a superconducting resonator. Moreover, $g_{s,t}$ does have a strong dependence on the QD confinement in the form of $1/\omega_d^2$.
Thus when the QD is larger, for example when $\hbar\omega_d=1$ meV, the coupling strength increases to $10^6$ s$^{-1}$ at $B_0=0.1$ T (about 20 times faster than the case when $\hbar\omega_d=8$ meV. Here, $d_0\sim \sqrt{\hbar/(m\omega_d)}$ is also increased for a larger dot.)
In short, benchmarked against spin dephasing rate, the strong coupling limit can be achieved more easily at lower magnetic field with s-SOC than with i-SOC.

Beside the transverse coupling, s-SOC also allows longitudinal coupling between a spin qubit and a superconducting resonator. The longitudinal coupling $g_{s,l}$ is in general finite in contrast to the vanishing $g_{s,l,isoc}$ in the case of the i-SOC. The magnitude of $g_{s,l}/g_{s,t}$ depends on the ratio of longitudinal and transverse gradient of magnetic field, with $g_{s,l}/g_{s,t}=b_{1l}/b_{1t}$. 
To make use of the longitudinal component, the spin quantization axis can also be adjusted to enhance the coupling strength of the s-SOC induced spin-photon coupling. 
\section{Discussion and Conclusions}
Our results in this study clearly demonstrate that spin decoherence, including relaxation and dephasing, can be modified strongly by the symmetry property of the interaction Hamiltonian under time-reversal operation.
\added{Conversely, the $B$-field dependence of spin relaxation or spin dephasing represents a hallmark to characterize symmetry properties of the SOC Hamiltonian in a system, which may serve as a tool to investigate the possible origin of decoherence in the system. 
In addition, because of the broken $\mathcal{T}$-symmetry of s-SOC, the property of spin-photon coupling can be modified (supplementary material), similar to the case of EDSR. The spin-photon transverse coupling strength shows a weaker $B_0$ dependence compared with the case of the i-SOC. The EDSR  and spin-photon coupling via s-SOC could also be further optimized through proper orientation of the applied magnetic field.}


\added{Our theory on the consequences of the $\mathcal{T}$-symmetry is valid when $||H_{SOC}||\ll E_Z \ll E_d$, which is satisfied in many experiments, with the purpose of protecting the spin qubit from electrical noise by reducing the spin-orbit mixing.}  Such a unified and general understanding of spin relaxation, pure dephasing, and spin control, and their connection to the $\mathcal{T}$-symmetry enables us to predict properties in many other physical systems, 
such as an electron spin in a double or triple QDs with or without a micromagnet. It can also be applied to the case of an electron spin qubit in a hybrid donor-QD \cite{tosi_silicon_2017} 
or a double donor system with different numbers of nuclei in each donor, where the hyperfine coupling difference could provide a longitudinal magnetic field gradient. Furthermore, in a double QD, a position dependent g-factor of an electron spin could arise due to the position dependent SOC (from inhomogeneous electric field) \cite{jock_silicon_2018, tanttu_controlling_2019}, and the g-factor difference in a double QD also results in an effective SOC that breaks the $\mathcal{T}$-symmetry. 
In short, the results obtained in this study, especially on how spin relaxation, dephasing, EDSR, and spin-photon coupling depend on the magnetic field amplitude $B_0$, field gradient $b_{1t}$ and $b_{1l}$, and field orientation $\theta$, are quite general.  They are the consequence of the time-reversal symmetry property of the given SOC, and are independent of the form of the orbital Hamiltonian.  As such they are valid whether in a single or double QD, or whether the spin mixing comes from spin-valley or spin-orbit coupling. 

In conclusion, we have studied spin decoherence and control in a single QD via an s-SOC generated by a micromagnet.
Based on symmetry analysis, we show that longitudinal effective field via the $\mathcal{T}$-symmetric i-SOC would vanish to the first order of SOC and all orders of Zeeman interaction, which we term as the longitudinal spin-orbit field cancellation. The absence of such a longitudinal field means the vanishing of pure dephasing for the electron spin.
On the other hand, we find that the s-SOC breaks the $\mathcal{T}$-symmetry, resulting in the elimination of both the van Vleck cancellation and the longitudinal spin-orbit field cancellation, and leading to an effective magnetic field that is different in form from the case of the i-SOC.
In particular, longitudinal effective field is allowed by s-SOC, while the transverse effective field does not depend on the applied field $B_0$. 
\added{Consequently, spin relaxation, pure dephasing, EDSR, and spin-photon coupling are all modified compared with the case of the i-SOC.}
Moreover, we show that s-SOC mediated EDSR could be optimized through through the orientation of the applied magnetic field.
%
Overall, our results clarify the decoherence and control properties of a spin qubit in the presence of an s-SOC in a single QD, and clearly reveal the underlying connection between the symmetry of the SOC and aforementioned spin properties.  These understandings could contribute significantly to our effort in building a scalable semiconductor quantum computer.

\section{Acknowledgment}
We thank the useful discussions with Jason Petta, David Zajac, Anthony Sigillito, and Felix Borjans on the spin relaxation experiments.
%
%
%
P.H. acknowledges supports by the National Natural Science Foundation of China (No. 11904157), the Science, Technology and Innovation Commission of Shenzhen Municipality (No. ZDSYS20170303165926217, No. JCYJ20170412152620376) and Guangdong Innovative and Entrepreneurial Research Team Program (Grant No. 2016ZT06D348); X.H. acknowledges support by US ARO via grant W911NF1710257.

\appendix

\section{Method}

To perform the SW transformation, the explicit form the generator can be obtained. From the model Hamiltonian and the definition of $L_d$, we have
$L_d^{-1}p_x=\frac{im^*}{\hbar}x$ and $L_d^{-1} x = \frac{-i}{\hbar m^{*}\omega_d^2}(p_x + m^*\omega_c y)$,
where $\omega_c=eB_{0z}/(m^*c)$ is the cyclotron frequency.
%
To the lowest order of $H_{s-SOC}$ and $E_Z/E_d$,
the generator $S$ is
\beq
S \approx L_d^{-1} H_{SOC} = \vec{\sigma}\cdot \vec{\eta},
\eeq
where $\vec{\eta}=\frac{-ig\mu_B\vec{b}_{1}}{2\hbar m^{*}\omega_d^2}(p_x + m^*\omega_c y)$.
Then, the noise term becomes
$\bra{\psi}[S,V_{ext}(\vec{r})]\ket{\psi}=-i\hbar \nabla_{p}{S}\cdot \nabla V_{ext}=\frac{1}{2}g\mu_B \vec{\sigma}\cdot \vec\Omega$,
where the electron is assumed to be in the ground orbital state $\ket{\psi}$.
\textbf{$\mathcal{T}$-symmetry of SOC Hamiltonian---}
Here, properties of the SOC under time-reversal are discussed in detail. To be more specific, we denote the Hamiltonian of the synthetic and the intrinsic SOC as $H_{s-SOC}$ and $H_{i-SOC}$.

For the i-SOC, it can be shown that $L_d^{-1}H_{i-SOC}\propto i\vec{\xi}(x,y)\cdot\vec\sigma$, where $\vec\xi(x,y)$ is function of position operators \cite{golovach2004, huang2013}. Thus, $L_d^{-1}H_{i-SOC}$ is TRS,
\beq
T_R L_d^{-1}H_{i-SOC} T_R^{-1} = L_d^{-1}H_{i-SOC}.
\eeq
where $T_R$ is the time-reversal operator (Note $T_R \vec{r} T_R^{-1} = \vec{r}$, $T_R i T_R^{-1} = - i$, and $T_R \vec{\sigma} T_R^{-1} = - \vec{\sigma}$).
In contrast, for the s-SOC,
\beq
T_R L_d^{-1}H_{s-SOC} T_R^{-1} \ne  L_d^{-1}H_{s-SOC},
\eeq
where $L_d^{-1}H_{s-SOC}$ contains TRA term.

Heuristically, these symmetry properties can simply be inferred if we consider an in-plane magnetic field, where the 2D vector potential $\vec{A}(\vec{r})$ vanishes. In this case, $\vec{p}$ is odd under TR: $T_R \vec{p} T_R^{-1} = - \vec{p}$. Given that $H_{i-SOC}\sim p_i\sigma_j$, thus, $H_{i-SOC}$ is TRS. Consequently, $L_d^{-1}H_{i-SOC}$ is also TRS since $H_d$ is TRS; In contrast, since $H_{s-SOC}\sim r_i\sigma_j$, we have $H_{s-SOC}$ and $L_d^{-1}H_{s-SOC}$ to be TRA.


\textbf{Parameters---}
For the numeric evaluation, the following values of parameters are used if not specified. We choose g = 2, $m^* = 0.19m_0$, and $\hbar\omega_d = 8$ meV for the effective g-factor, the effective mass, and the orbital confinement of an electron in a silicon QD.
For the SOC constants, we choose the Rashba constant as $\alpha=10$ m/s, and the Dresselhaus constant is set to zero for rough estimation.
We choose $b_{1t}=b_{1l}=0.1$ mT/nm for the transverse and longitudinal field gradient \cite{yoneda2018}.
We choose $v_1=5900$ m/s and $v_2=v_3=3750$ m/s for the speed of the different acoustic phonon branches, $\rho_c=2200$ kg/m$^3$ for the mass density,
$\Xi_d=5.0$ eV and $\Xi_u=8.77$ eV for the dilation and shear deformation potential constants.
The electron temperature is set to be zero for simplicity. 
%
%
We choose the amplitude $\sqrt{A}=5$ $\mu eV$ for the $1/f$ charge noise, the cutoff frequency $\omega_0$=1 $s^{-1}$, and length scale $l_0=100$ nm.
For the s-SOC induced EDSR driving of spin qubit, we choose the following parameters if not specified. We choose the maximum electric field $E_{max}=6\times10^5$ $V/m$ for the microwave driving.
In silicon QD, the valley physics can modify spin relaxation at low magnetic field via the spin-valley mixing \cite{yang2013, tahan2014, huang2014b, hao_electron_2014}. 
{Here, we assume the Zeeman splitting is away from the valley splitting so that the spin-valley mixing is suppressed and we focus on the intra-valley spin-orbit mixing.} 



\section{Spectral density of electrical noise}

In the section, we present the detailed derivation of the spectral density of phonon noise and Johnson noise. 

\textbf{Electron-phonon interaction---}
The electron-phonon interaction $H_{EP}$ is one of the noise source provides the energy dissipation. In silicon QD, we have \cite{yu_fundamentals_2010} 
\begin{equation}
H_{EP}=\sum_{\vec{q}j}f(q_z)e^{i\vec{q}\cdot\vec{r}} M_{\vec{q}j}^{(z)} (b_{-\vec{q}j}^\dag + b_{\vec{q}j}), \label{Hep}
\end{equation}
\beq
M_{\vec{q}j}^{(z)} =i\sqrt{\hbar /(2\rho_c qv_{j})} q\Xi_{\vec{q}j}^{(z)}, \label{Mqj}
\eeq
\beqa
{\Xi}_{\vec{q}j}^{(z)}
&=& \hat{e}^{(\vec{q}j)} \cdot (\Xi_{d}\tensor{1}+ \Xi_{u}\tensor{U}^{(z)}) \cdot \hat{q}, \label{Xiqj}
\eeqa
where
$b_{\vec{q}j}^\dag$ ($b_{\vec{q}j}$) is the creation (annihilation) operator of a phonon with wave vector $\vec{q}$ and branch-index $j$, $j=l$ (longitudinal mode), $t_1$, or $t_2$ (transverse modes).
The function $f(q_z)=\bra{\psi_z(z)}e^{iq_zz}\ket{\psi_z(z)}\sim e^{-q^2d_z^2/4}$ equals unity for $|q_z| \ll d_z^{-1}$, and vanishes for $|q_z|\gg d_z^{-1}$, where $d_z$ is the characteristic size of the quantum well along the z axis.
$\rho_c$ is the sample density, $v_j$ is phonon velocity, $\hat{e}^{(\vec{q}j)}$ and $\hat{q}$ are unit vectors of phonon polarization and  wave vector, $\Xi_{d}$ and $\Xi_{u}$ are the dilation and uniaxial shear deformation potential constants. Each component of  $\Xi_{\vec{q}j}^{(z)}$ can be evaluated as
$\Xi_{\vec{q}l}=\Xi_d+\Xi_u\cos^2\theta$ (LA), $\Xi_{\vec{q}t_1}=0$, and $\Xi_{\vec{q}t_2}=\Xi_u\cos\theta\sin\theta$ (TA), where $\theta$ is polar angle of phonon wave-vector with respect to the growth direction (crystal axis [001]).

\textbf{Spectral density of phonon noise---}
The correlation of the electric force due to phonons, $F=-eE(\vec{r})=-\vec{\nabla}U_\mathrm{ph}(\vec{r})$, is thus given by (x component),
\begin{eqnarray}
\lefteqn{\left\langle F_x(0)F_x(t)\right\rangle = \sum_{\boldsymbol{q}j}\frac{\left|f(q_z)\right|^2} {{2\rho_c\omega_{qj}/\hbar}} q_x^2e^{i\vec{q}_\parallel\cdot\vec{r}} }\nonumber\\
&&\times  \left|q\Xi_{\boldsymbol{q}j}\right|^2(b_{\boldsymbol{q}j}b_{\boldsymbol{q}j}^\dag e^{i\omega_{qj}t}+b_{-\boldsymbol{q}j}^\dag b_{-\boldsymbol{q}j}e^{-i\omega_{qj}t}). \label{ExEx}
\end{eqnarray}
We consider the adiabatic condition, where the energy scale of the noise is much less than the dot confinement energy $E_d=\hbar\omega_d$ and the valley splitting, so that the electron orbital state stays in the instantaneous ground state
$
\psi (\vec{r})=\exp\left( -(\vec{r}-\vec{r}_0)^2/2\lambda ^{2}\right) /\lambda \sqrt{\pi },
$
where $\lambda ^{-2}=\hbar ^{-1}\sqrt{(m^{\ast }\omega _{d})^{2}+(eB_{z}/2c)^{2}}$ is the effective radius. Then, we simplify the exponential terms $e^{i\vec{q}_\parallel\cdot\vec{r}}$ by its mean-field value
$e^{-q_\parallel^2\lambda^2/4}$.

The spectrum of the phonon noise in the $x$-direction is 
\beqa
\lefteqn{S_{FF}(\omega)=\mathrm{Re}\frac{1}{2\pi}\int_{-\infty}^{\infty}dt\, \langle F_{x} F_{x}(t)\rangle \cos(\omega t)} \nonumber\\
&=& \sum_{j}\frac{\hbar\omega^5(2N_{\omega}+1)}{16\pi^2\rho_cv_j^7}\int_0^{\pi/2} d\theta \bar{\Xi}_{j\theta}^2\sin^3\theta f_j(\omega,\theta),
\nonumber
\eeqa
where $N_{\omega}=[\exp(\hbar\omega/k_BT)-1]^{-1}$ is the phonon excitation number and the cutoff function $f_j(\omega,\theta) = \left|f\left({\omega}\cos\theta/{v_j}\right)\right|^2e^{-\omega^2\lambda^2\sin^2\theta/2v_j^2}$ is due to
the suppression of the matrix element for the electron-phonon interaction in a large QD. 
For the noise in the $y$ and $z$ direction, we have $S_{FF,y}(\omega)= S_{FF}(\omega)$ and
\beqa
\lefteqn{S_{FF,z}(\omega)=\sum_{j}\frac{\hbar\omega^5(2N_{\omega}+1)}{8\pi^2\rho_cv_j^7}}\nonumber\\
&&\times\int_0^{\pi/2} d\theta \bar{\Xi}_{j\theta}^2\sin\theta\cos^2\theta f_j(\omega,\theta).
\nonumber
\eeqa
If the dipole approximation $e^{i\vec{q}_\parallel\cdot\vec{r}}\approx 1+ i\vec{q}_\parallel\cdot\vec{r}$ is employed (valid when the magnetic field is weak), then, we have $f_j(\omega,\theta)=1$.  Furthermore, the temperature $T$ of the lattice vibration is normally very low ($T<1$ K), so that $2N_{\omega}+1=\coth(\hbar\omega/2k_{B}T)\approx 1$, in which case the spectrum of phonon noise shows a nice $\omega^5$ dependence.

\textbf{Interaction of an electron to Johnson noise or $1/f$ charge noise---}
For the Johnson noise or $1/f$ charge noise, the photon wave vector is small due to the fast speed of light. Thus, the dipole approximation is applicable,
\beq
V_{ext}(\vec{r})=-\vec{r}\cdot \vec{F}(t),
\eeq
where $\vec{F}(t)$ is the random force due to the Johnson noise or $1/f$ charge noise.

\textbf{Spectral density of Johnson noise---}
The spectrum of Johnson noise is given by \cite{huang2014b} 
\begin{equation}
S_{V}\left( \omega \right) =2\xi \omega \hbar ^{2}f_c(\omega_Z)\coth \left( \hbar \omega /2k_{B}T\right), \label{Sv_John}
\end{equation}
where $S_V$ is the spectrum of electrical voltage $S_{V}(\omega)=\frac{1}{2\pi }\int_{-\infty }^{+\infty }\overline{V\left( 0\right)  V\left( t\right) } \cos \left( \omega t\right) dt$, $\xi =R/R_{k}$ is a dimensionless constant, $R_{k}=h/e^{2}=26$ k$\Omega $ is the quantum resistance, and $R$ is the resistance of the circuit. $f_c(\omega)=1/[1+\left( \omega/\omega _{R}\right) ^{2}]$ is a natural cutoff function for Johnson noise, where $\omega _{R}=1/RC$ is the cutoff frequency, and $C$ is capacitors in parallel with the resistance $R$.

The Johnson noise of the circuits outside the dilution refrigerator is generally well-filtered. Thus we consider only Johnson noise of the low-temperature circuit inside a dilution refrigerator. The corresponding spectrum of electrical force is 
$S_{FF,i} \left( \omega \right) = S_{V} (\omega)/l_{0}^{2}$, 
where $i=X,Y$, or $Z$, and $l_{0}$ is the length scale of the QD.  

\end{document}